\documentclass[aps,preprint,pra]{revtex4-1}
\usepackage{bm}
\usepackage{bbm}
\usepackage{subfigure}
\usepackage{natbib}
\usepackage{graphicx}
\usepackage{verbatim}
\usepackage{amssymb}
\usepackage{amsmath}
\usepackage{amsfonts}
\usepackage{dsfont}

\newcommand{\nsp}{\hspace{-0.4pt}}
\newcommand{\ssp}{\hspace{0.4pt}}

\newcommand{\be}{\begin{equation}}
\newcommand{\ee}{\end{equation}}
\newcommand{\bea}{\begin{eqnarray}}
\newcommand{\eea}{\end{eqnarray}}

\newcommand{\commut}[2]{[\ssp #1\ssp,\,#2\ssp]}
\newcommand{\commutb}[2]{\big[\ssp #1\ssp,\,#2\ssp\big]}

\newcommand{\ket}[1]{|#1\rangle}

\newcommand{\bk}[1]{\langle#1\rangle}

\newcommand{\one}{1 \hspace{-1.0mm}  {\bf l}}
\newcommand{\Tr}{\text{Tr}}
\newcommand{\tr}{\text{tr}}

\newcommand{\Cov}{\text{Cov}}
\newcommand{\Ad}{\text{Ad}}
\newcommand{\Pf}{\text{Pf}}

\DeclareRobustCommand\openzero{\leavevmode\hbox{0\kern-.55em0}}

\newcommand{\bibliopath}{/Users/carollo/GoogleDrive/Documents/Uni/Articles&Manuscripts/}

\begin{document}

\title{Symmetric Logarithmic Derivative of Fermionic Gaussian States}
\author{ Angelo Carollo$^{1,2}$, Bernardo Spagnolo$^{1,2,3}$ and Davide Valenti$^{1,4}$}

\affiliation{$^1$Dipartimento di Fisica e Chimica, Group of
Interdisciplinary Theoretical Physics and CNISM, Universit\`{a} di
Palermo, Viale delle Scienze, Edificio 18, I-90128
Palermo, Italy\\
$^2$Radiophysics Department, Lobachevsky State University of Nizhni Novgorod, 23 Gagarin Avenue, Nizhni Novgorod 603950, Russia\\
$^3$Istituto Nazionale di Fisica Nucleare, Sezione di Catania, Via
S. Sofia 64, I-90123 Catania, Italy\\
$^4$IBIM-CNR Istituto di Biomedicina ed Immunologia Molecolare
``Alberto Monroy", Via Ugo La Malfa 153, I-90146 Palermo, Italy
 }



\begin{abstract}
In this article we derive a closed form expression for the symmetric logarithmic derivative of Fermionic Gaussian states. This provides a direct way of computing the quantum Fisher Information for Fermionic Gaussian states. Applications ranges from quantum Metrology with thermal states and non-equilibrium steady states with Fermionic many-body systems.
\end{abstract}
\maketitle







\section{Introduction}
Quantum metrology or quantum parameter estimation is the theory that studies the accuracy by which a physical parameter of a quantum system can be estimated through measurements and statistical inference. 
In many physical scenarios, quantities which are to be estimated may not be directly
observable, either due to experimental limitations or on account of fundamental principles. When this is the case, one needs to infer the value of the variable after measurements on a given probe. This is essentially a parameter
estimation problem whose solution may be found using methods from
classical estimation theory \cite{Cramer1946} or, when quantum systems are
involved, from its quantum counterpart \cite{Helstrom1976}. 
Quantum metrology finds applications in a range of diverse fields, from fundamental physics, such as improving frequency and time standards~\cite{Udem2002,Katori2011,Giovannetti2004}, estimating parameters in quantum field theory~\cite{Aspachs2010,Ahmadi2014},  improving the accuracy of gravitational wave interferometry~\cite{Schnabel2010,Aasi2013}, to applied physics, such as thermometry~\cite{Correa2015,DePasquale2016}, spectroscopy~\cite{Schmitt2017,Boss2017}, imaging~\cite{Tsang2016,Nair2016,Lupo2016}, magnetic field detection~\cite{Taylor2016,Bonato2016} navigation~\cite{Cai2013,Komar2014} and remote sensing~\cite{Dowling2015,Boto2000}. Exploiting the quantum nature of physical systems provides remarkable advantage in enhancing the accuracy of estimation problems, and exploring this possibility plays a pivotal role in the current swift development of quantum technology~\cite{Caves1981,Huelga1997,Giovannetti2006,Paris2009,Giovannetti2011,Toth2014,Szczykulska2016,Pezze2016,Nichols2017,Braun2017}.
Simultaneous quantum estimation of multiple parameters provides better precision over individual estimation strategies with equivalent resources~\cite{Humphreys2013,Baumgratz2016}. This fact has sparked interest in multi-parameter quantum metrology in a variety of scenarios~\cite{Humphreys2013,Baumgratz2016,Pezze2017,Apellaniz2018}.

Recent advances in quantum metrology have shown that the accuracy in parameter estimation can be enhanced by employing peculiar quantum many-body state as a probe~\cite{Zanardi2008,Braun2017}. Conversely, quantum metrology tools may well be exploited in the characterisation of many-body systems. Noteworthy instances of many-body quantum systems are those experiencing quantum phase transitions. Indeed, quantum parameter estimation, with its intimate relation with geometric information, provides a novel and promising approach in the characterisation of equilibrium~\cite{Carollo2005,Zhu2006,Hamma2006,Zanardi2006,CamposVenuti2007,CamposVenuti2008,Zanardi2007,Zanardi2007a,Garnerone2009a,Rezakhani2010} and out-of-equilibrium quantum critical phenomena~\cite{Banchi2014,Marzolino2014,Kolodrubetz2016,Carollo2017,Marzolino2017}. 

Fermionic Gaussian states play a major role in the derivation of exact and approximate solution of many-body problems of Fermionic and spin systems. Deriving closed form expressions of quantities involved in parameter estimation problems for many body quantum system is a major challenge. This work addresses this task in the special, yet relevant, case of arbitrary Fermionic Gaussian states.

The solution of a parameter estimation problem amounts to find
an estimator, {\em i.e} a mapping $\hat{\bm{\lambda}}=\hat{\bm{\lambda}}
(x_1,x_2,...)$ from the set $\chi$ of measurement outcomes into
the space of parameters $\bm\lambda \in\mathcal{M}$.  Optimal estimators in classical
estimation theory are those saturating the Cramer-Rao (CR) inequality,
\be\label{eq:CCRB}
\Cov_{\bm\lambda}[\hat{\bm{\lambda}}] \geq J^{c} (\bm \lambda)^{-1} 
\ee
which poses
a lower bound on the mean square error $\Cov_{\lambda}
[\hat{\bm{\lambda}}]_{\mu\nu} = E_{\lambda} [(\hat{\lambda} -
\lambda)_\mu(\hat{\lambda}-\lambda)_\nu]$ in terms of the Fisher
information (FI)
\be
J^{c}_{\mu\nu}(\bm\lambda) = \int_\chi d\hat{\bm{\lambda}}(x)\,
p(\hat{\bm{\lambda}}|\lambda) \partial_\mu \log
p(\hat{\bm{\lambda}}|\lambda) \partial_\nu \log
p(\hat{\bm{\lambda}}|\lambda)\:.
\ee
 For unbiased estimators, the mean square error is equal to the
covariance matrix $\Cov_{\bm\lambda} [\hat{\bm{\lambda}}]_{\mu\nu} = E_{\bm\lambda}
[\hat{\bm{\lambda}}_\mu\hat{\bm{\lambda}}_\nu]  - E_{\bm\lambda} [\hat{\bm{\lambda}}_\mu]
E_{\bm\lambda}[\hat{\bm{\lambda}}_\nu]\:.$
The expression~(\ref{eq:CCRB}) should be understood as a matrix inequality. In general, one writes 
\[
\tr(W\Cov_{\bm\lambda}[\hat{\bm{\lambda}}] )\ge\tr(W J^{c}(\bm\lambda)^{-1}),
\]
where $W$ is a given positive definite cost matrix, which allows the uncertainty cost of different parameters to be weighed unevenly.

In the classical estimation problem, both in the single parameter case, and in the multi-parameter one, the bound is saturable in the limit of an infinite number of repetitions of an experiment using the maximum likelihood estimator \cite{Kay1993}.
However, an interesting difference between multi-parameter and single parameter metrology arises due to the correlation between parameters. Indeed, it may well happen that the resulting Fisher information matrix is non-diagonal. This means that the estimators for the parameters will not be independent. In a separate scheme in which all parameters except the $\lambda_{\mu}$ are perfectly known, the single parameter CR bound implies that the uncertainty of estimating $\lambda_{\mu}$
 is lower bounded by $\text{Var}(\hat{\bm\lambda}) \geq 1/J^{c}_{\mu\mu}$. On the other hand, in the simultaneous scenario in which all parameters are estimated at the same time, one finds $\text{Var}(\hat{\bm\lambda}) \geq (J^{c}(\bm\lambda)^{-1})_{\mu\mu}$. From  basic algebra of  positive-definite matrices, we have that
$(J^{c}(\bm\lambda)^{-1})_{\mu\mu} \geq 1/J^{c}(\bm\lambda)_{\mu\mu}$, with equality holding only in the case when all off-diagonal elements vanish.
Since asymptotically the CR bound is saturable, it implies that the equivalence between the simultaneous and separate scheme in the limit of a large number of experiment repetitions can only hold if $F$ is a diagonal matrix, and hence there are no statistical correlations between the estimators \cite{Cox1987}.

Clearly, for any real positive definite matrix one can perform an orthogonal rotation to a new basis in which the matrix is diagonal. This simply means that there are always linear combinations of the parameters for which the diagonality conditions hold. This choice should be, however, contrasted with the physical opportunity of performing such rotation, as the choice of the parameters we are interested in may arise as a result of physical considerations and in this sense determine a preference in a specific basis.

While the fundamental objects in classical Fisher information are parameter-dependent probability-distribution of the data, the fundamental objects involved in the quantum estimation problem are the density matrices $\rho(\bm\lambda)$  labelled by ${\bm\lambda\cal \in M}$. In the quantum scenario we therefore face an additional challenge of determining the optimal measurement for extracting most of the information on the parameters of interest from the quantum states. In the single parameter case the
situation is relatively simple. Maximization of the classical Fisher information over all quantum measurements yields the quantity referred to as the quantum Fisher information (QFI). The key object involved in the calculation of the QFI is the so called \emph{symmetric logarithmic derivative} (SLD), $L$, which is implicitly defined as the Hermitian operator satisfying the
equation
\be\label{eq:SLD}
\frac{d \rho}{d\lambda}
= \frac12 \left ( \rho  L +
L \rho  \right )\:.
\ee
The QFI can be calculated using the formula:
\be
 J =\Tr (\rho L^{2}),
 \ee
One can always choose the projective measurement in the eigenbasis of the SLD which yields FI equal to the QFI. Hence, the QFI determines the ultimate achievable precision of estimating the parameter on density matrices $\rho(\bm\lambda)$ in the asymptotic limit of an infinite number of experiment repetitions.
In a multiparameter scenario, a direct generalization of single parameter CR bound leads to the multiparameter QFI CR bound~\cite{Helstrom1976,Holevo2011,Paris2009}, that reads
\begin{equation}\label{eq:SCRB}
\tr(W \Cov(\hat{\bm{\lambda}}))\ge\tr(W J^{-1}),
\end{equation}
 where 
 \be
 J_{\mu\nu}=\frac{1}{2}\Tr\rho\{L_{\mu},L_{\nu}\},
 \ee
is the quantum Fisher information matrix (QFIM), $W$ is the cost matrix, and $L_{\mu}$ is the SLD implicitly defined by~(\ref{eq:SLD}), with $\rho$ derived with respect to the parameter $\lambda_{\mu}$.\\
Unlike the single parameter case, in the multi-parameter scenario the QFI CR bound cannot always be saturated. Intuitively, this is due to the incompatibility of the optimal measurements for different parameters. A sufficient condition for the saturation is indeed $[L_{\mu},L_{\nu}]=0$, which is however not a necessary condition. Within the comprehensive framework of quantum local asymptotic normality (QLAN)~\cite{Hayashi2008,Kahn2009,Gill2013,Yamagata2013}, a necessary and sufficient condition for the saturation of the multi-parameter CRB is given by~\cite{Ragy2016} 
\be
\mathcal{U}_{\mu\nu}=-\frac{i}{4}\Tr\rho [L_{\mu},L_{\nu} ] = 0 \qquad\forall \mu,\nu, 
\ee
and it is known as \emph{compatibility condition}~\cite{Ragy2016}. In the context of quantum information geometry, and quantum holonomies of mixed states, $\mathcal{U}_{\mu\nu}$ is known as mean Uhlmann curvature~\cite{Carollo2017}.\\
In this paper, we derive a closed form expression of the SLD of Fermionic Gaussian states, which are of fundamental importance in the analysis of steady-states of both equilibrium and non-equilibrium quantum many-body systems, and their applications to quantum metrology. \\
The paper is organised as follows. In the next section we shortly review the main properties of Fermionic Gaussian states. In section 3, an explicit form for the calculation of the SLD is derived. In the last section, the conclusions are drawn.

\section{Fermionic Gaussian State}

We review here the main properties of Fermionic Gaussian states (FGSs). Let's consider a systems of $n$ fermionic particles described by creation and annihilation operators $c_{j}^{\dag}$ and $c_{j}$. These operators obey the canonical anticommutation relations,
\be
\{c_{j},c_{k}\}=0\qquad \{c_{j},c^{\dag}_{k}\}=\delta_{jk}\,.
\ee
Let's define the Hermitian Majorana operators as
\be
\omega_{2j-1}:=c_{j}+c^{\dagger}_{j}\,,\qquad \omega_{2j}:=i(c_{j}-c_{j}^{\dagger})\,,
\ee
which are generators of a Clifford algebra, and satisfy the following anti-commutation relations
\be
\{\omega_{j},\omega_{k}\}=2\delta_{jk}\,.
\ee
Fermionic Gaussian states are defined as states that can be expressed as
 \begin{equation}\label{GS}
\rho=\frac{e^{-\frac{i}{4}\bm{\omega}^{T} \Omega \bm{\omega}}}{Z}\,,\qquad Z:=\Tr[ e^{-\frac{i}{4}\bm{\omega}^{T} \Omega \bm{\omega}}]
 \end{equation}
 where $\Omega$ is a $2n\times 2n$ real antisymmetric matrix and $\bm \omega:=(\omega_{1}\dots \omega_{2n})^{T}$ is a 2n-dimensional array of Majorana Fermions. 
As any antisymmetric real matrix, $\Omega$ can be cast in the following canonical form by an orthogonal matrix $Q$,
i.e.
\begin{align}
\Omega &= Q^T\, \bigoplus_{k=1}^n
\begin{pmatrix} 0&\Omega_k\\-\Omega_k&0 \end{pmatrix}
  \, Q & Q^T=Q^{-1}~,
  \label{e.Q}
\end{align}
where $\pm i\Omega_k$ are $\Omega$'s  eigenvalues. Let
\be
\bm{z}=(z_{1},\dots,z_{2n})^{T}:=Q\bm{\omega}
\ee
be the vector of Majorana fermions in the eigenmode representation. Hence,
\begin{align}
\rho &= \frac1Z\prod_k \left[\cosh\left(\frac{\Omega_k}2\right) -i
\sinh\left(\frac{\Omega_k}2\right)\,z_{2k-1} z_{2k}\right]~,
\\
Z &= \prod_k 2\cosh\left(\frac{\Omega_k}2\right) \,.
\label{e.gaussianZ}
\end{align}

 Gaussian states are completely specified by the two-point correlation matrix 
 \be
 \Gamma_{jk}:=1/2\Tr{(\rho[\omega_{j},\omega_{k}])}\,, \qquad \Gamma=\Gamma^{\dagger}=-\Gamma^{T}\,,
 \ee
 which is an imaginary antisymmetric matrix. Let's recall some basic properties of the correlation function. As for any Fermionic States, all odd-order correlation functions of FGS are zero, due to the parity super-selection rule. In FGS, all even-order correlations, higher than two, can be obtained from  $\Gamma$ by Wick's theorem~\cite{Bach1994} , i.e. 
\be
\Tr(\rho \omega_{k_{1}} \omega_{k_{2}}...\omega_{k_{2p}})=\Pf(\Gamma_{k_{1}k_{2}\dots k_{2p}}), \qquad  1 \le k_{1} < . . . < k_{2p} \le 2n
\ee
and $\Gamma_{k_{1}k_{2}\dots k_{2p}}$ is the corresponding $2p \times 2p$ submatrix of $\Gamma$. $\Pf(\Gamma_{k_{1}k_{2}\dots k_{2p}})^{2} = \det (\Gamma_{k_{1}k_{2}\dots k_{2p}})$ is the Pfaffian. 
An especially useful case is the four-point correlation function
\begin{equation}\label{Pf}
	\Tr{(\rho\omega_{j} \omega_{k}\omega_{l}\omega_{m})}= a_{jk}a_{lm}-a_{jl}a_{km}+a_{jm}a_{kl},
\end{equation}    
where $a_{jk}:=\Gamma_{jk}+\delta_{jk}$.  As 
 \be
 \Gamma_{jk}=\frac{2i}Z\frac{\partial Z}{\partial \Omega_{jk}}
 \ee 
 one can show that
\begin{align}
\Gamma = \tanh\left(i\frac{\Omega}2\right)~.
\label{e.corrG}
\end{align}
The correlation matrix is diagonal in the same basis of $\Omega$
and its eigenvalues read $\gamma_k = \tanh(\Omega_k/2)$. Hence
\begin{align}
  \rho &= \prod_{k=1}^{n} \frac{1-i |\gamma_k|\,z_{2k-1} z_{2k}}2~,
  \label{e.gaussianC}
\end{align}
where $|\gamma_k|\le 1$. Hence the Gaussian fermionic state can be factorised into a tensor product $\rho=\bigotimes_{k}\rho_{k}$ of density matrices of the eigenmodes $\rho_{k}:=\frac{1-i |\gamma_k |\,z_{2k-1} z_{2k}}2$.
Note that for $\gamma_k=\pm 1$, one has $\Omega_k = \pm\infty$, making the definition~(\ref{GS}) of Gaussian state not well defined, unlike Eq.~\eqref{e.gaussianC}, showing that the latter offer an appropriate parameterisation even in those extremal points. Notice that $|\gamma_{k}|=1$ corresponds to a fermionic mode $\tilde{c}_{k}=1/2(z_{2k-1}+z_{2k})$ being in a pure state, as  it is clear from the following
explicit expression for
the purity of the states $\rho_{k}$:
\begin{align}
  \Tr[\rho_{k}^2] =
  \frac{\det\left[2\cosh\left(\,\Omega_{k}\right)\right]^{\frac12}}{
  \det\left[2\cosh\left(\frac{\Omega_{k}}2\right)\right]} ~.
  \label{e.purity}
\end{align}

\begin{align}
  \Tr[\rho^2] =
  \frac{\det\left[2\cosh\left(i\,\Omega\right)\right]^{\frac12}}{
  \det\left[2\cosh\left(i\frac{\Omega}2\right)\right]} =
  \sqrt{\det\left(\frac{1+\Gamma^2}2\right)}~.
  \label{e.purity1}
\end{align}

\section{Symmetric Logarithmic Derivative of Fermionic Gaussian States}
We will derive here an explicit formula for the calculation of the SLD for Fermionic Gaussian states. To this end we review a useful expression adapted from reference~\cite{Jiang2014} needed for the derivation of the symmetric logarithmic derivative of density matrices in the exponential form 
\be\label{eq:exp}
\rho=e^{D(\bm{\lambda})}.
\ee
Clearly, a Gaussian Fermionic state can be expressed in the exponential form~(\ref{eq:exp}) by identifying 
\be \label{eq:D}
D=-\frac{i}{4}\bm{\omega}^{T}\cdot \Omega\cdot\bm{\omega} - \one \ln{Z}.
\ee
Notice, that the above parameterisation is well defined only in the case of full-rank density matrices. As usual, the case of extremal conditions $|\gamma_{k}|=1$, where is an eigenvalue of the correlation function should be carried out as a limiting procedure.
 
The starting point is the expression derived in Eq.~(2.1) of Ref.~\cite{Wilcox1967} for derivative of density operators
\begin{align}\label{eq:wilcox}
 \dot  \rho = \int_0^1 e^{sD}\,\dot  D\, e^{(1-s)D}\,d s\;,
\end{align}
where dots represent derivatives with respect to a parameter $\lambda$.  One can use the nested-commutator relation
\begin{align}
e^D A e^{-D}&=A+[D,A]+\frac{1}{2!}\,\big[D,[D,A]\,\big]+\cdots\nonumber\\
&=\sum_{n=0}^\infty\frac{1}{n!}\,\mathcal{C}^n(A)=e^{\mathcal{C}}(A)\;,
\end{align}
where $\mathcal{C}^n(A)$, a linear operation on $A$, denotes the $n$th-order nested commutator $\commutb{D}{\nsp\ldots\,,\commut{D}{A}}$, with $\mathcal{C}^0(A)=A$.  Applying this relation to the expression~(\ref{eq:wilcox}) leads to
\begin{align}\label{eq:rhodot}
\begin{split}
 \dot  \rho \rho^{-1}
 &= \dot  D +\frac{1}{2!}\,\commut{D}{\dot  D}+\frac{1}{3!}\,\commutb{D}{\commut{D}
 {\dot D}}+\cdots\\
 &= \sum_{n=0}^\infty \frac{1}{(n+1)!}\, \mathcal{C}^n(\dot D)
 = h(\mathcal{C})(\dot D) \;,
\end{split}
\end{align}
where $h$ is the generating function of the expansion coefficients in Eq.~(\ref{eq:rhodot}),
\begin{align}\label{eq:generating_g}
 h(t) = 1+\frac{t}{2!}+\frac{t^2}{3!}+\cdots = \frac{e^t - 1}{t}\;.
\end{align}

Using the definition of symmetric logarithmic derivative, i.e. 
\be
\dot \rho = \frac{1}{2}\left( L\rho + \rho L\right)~,
\ee
and that of density matrix in exponential form~(\ref{eq:exp}), one gets 
\begin{align}\label{eq:left_derivative_b}
\begin{split}
\dot  \rho\, \rho^{-1}  &= \frac{1}{2}\big(L + e^D L e^{-D}\big)\\[2pt]
&= \frac{1}{2} \bigg(L + \sum_{n=0}^\infty \frac{1}{n!}\, \mathcal{C}^n(L) \bigg)
= r(\mathcal{C})(L)\;,
\end{split}
\end{align}
where the generating function is $r(t)=(e^t+1)/2$.  Suppose that the SLD adopts the form,
\begin{align}\label{eq:expansion_L_a}
L &= \sum_{n=0}^\infty f_n\, \mathcal{C}^n(\dot  D)
=f(\mathcal{C})(\dot D)\;,
\end{align}
with the generating function 
\begin{align}\label{eq:f_expansion_a}
f(t) = f_0+f_1 t+ f_2 t^2+\cdots 
\end{align}
to be determined.
Plugging Eq.~(\ref{eq:expansion_L_a}) into Eq.~(\ref{eq:left_derivative_b}) yields
\begin{align}\label{eq:combining_expansions}
 \dot  \rho\, \rho^{-1}= r(\mathcal{C})\bigl[f(\mathcal{C})(\dot D)\bigr] =r\circ f(\mathcal{C})(\dot D)=r \cdot f(\mathcal{C})(\dot D)~,
\end{align}
where the identity $r\circ f = r\cdot f $, between the combination, $r\circ f$, of the two functions and their simple product, $r\cdot f$,  arises from $\mathcal{C}^n(\mathcal{C}^m(A))=\mathcal{C}^{n+m}(A)$. Comparing Eq.~(\ref{eq:combining_expansions}) with Eq.~(\ref{eq:rhodot}) leads to the following relation between generating functions,
\begin{align}\label{eq:f_expansion_b}
 f(t)= \frac{h(t)}{r(t)}=\frac{\tanh(t/2)}{t/2}
 = \sum_{n=0}^\infty \frac{4\, (4^{n+1}-1) B_{2n+2}}{(2n+2)!}\,t^{2n}\;,
\end{align}
where $B_{2n+2}$ is the $(2n+2)$th Bernoulli number.  Comparing Eqs.~(\ref{eq:f_expansion_a}) with (\ref{eq:f_expansion_b}), we have
\begin{align}
 f_n =
\begin{cases}
 \displaystyle{\frac{4\, (4^{n/2+1}-1) B_{n+2} }{(n+2)!}}\,,&\mbox{for even $n$}\,,\\[3pt]
  0\,,&\mbox{for odd $n$}\,.
\end{cases}
\end{align}
The vanishing of the odd-order of $f_n$s is a consequence of the Hermiticity of $L$, which makes $f(t)$ an even function.\\
From the definition~(\ref{eq:D}) of $D$, one straightforwardly finds:
\be
\dot D=-\frac12\ssp\ssp {\bm \omega}^{T}\cdot\, {\dot \Omega}\cdot \bm \omega - \frac{\dot Z}{Z} \one \;,
\ee
which shows that $\dot D$ is itself a quadratic function of the Majorana Fermion operators, where $\dot\Omega$ is an antisymmetric real matrix given by
\begin{align}
\dot \Omega &= Q^T\, \bigoplus_{k=1}^n
\begin{pmatrix} 0&\dot \Omega_k\\-\dot \Omega_k&0 \end{pmatrix} Q + i [R, \Omega] ~,
  \label{e.dD}
\end{align}
where $R:= i Q^{T}\dot Q$.
Therefore, for a Gaussian Fermionic state the operator $\dot D$ can be written in a canonical form in terms of the Majorana Fermions of its own eigen-modes, as
\be 
\dot D=-\frac{i}{4}\sum_{k} \tilde{\Omega}_{k}[\tilde z_{2k-1}, \tilde z_{2k}] - \frac{\dot Z}{Z} \one =\sum_{k} \tilde{\Omega}_{k}\Big(\tilde c_{k}^{\dagger} \tilde c_{k}-\frac12\Big) - \frac{\dot Z}{Z} \one~.
\ee
where  $\tilde c_{k}:=\frac{1}{2}(\tilde z_{2k-1}+i\tilde z_{2k})$, $\tilde c_{k}^{\dagger}:=\frac{1}{2}(\tilde z_{2k-1}-i\tilde z_{2k})$ are the ordinary annihilation and creation operators of the eigen-modes of $\dot D$ and $\tilde \Omega_{k}$ the corresponding eigenvalues. 
It is straightforward to derive the commutation relations between $\dot D$ and Fermionic operators, 
\begin{align}\label{eq:commutation_relations}
 \commutb{\dot D}{\tilde c_k} =- \tilde{\Omega}_k \tilde c_k\,,\quad \commutb{\dot D}{\tilde c_k^\dagger} = \ssp \tilde{\Omega}_k \tilde c_k^\dagger\;,
\end{align}
and for quadratic operators, also
\begin{align}
\commutb{\dot D}{\tilde c_j^{\dagger}\tilde c_k} =(\tilde{\Omega}_j - \tilde{\Omega}_k) \tilde c_j^{\dagger}\ \tilde c_k,\quad \commutb{\dot D}{\tilde c_j^{\dag} \tilde c_k^{\dag}} =(\tilde{\Omega}_j + \tilde{\Omega}_k) \tilde c_j^{\dag} \tilde c_k^{\dag}.\end{align}
Consequently, one finds
\begin{align}
 f(\mathcal C)\ssp(\tilde c_j^\dagger \tilde c_k) = f(\tilde{\Omega}_j-\tilde{\Omega}_k)\ssp \tilde c_j^\dagger \tilde c_k\;,\\[2pt]
 f(\mathcal C)\ssp(\tilde c_j^{\dag} \tilde c_k^{\dag}) = f(\tilde{\Omega}_j+\tilde{\Omega}_k)\ssp \tilde c_j^{\dag} \tilde c_k^{\dag}\;.\label{eq:relation_b}
\end{align}
The above expression, plugged into formula~(\ref{eq:expansion_L_a}), shows that $L$ is at most quadratic in Fermionic operators.

Due to the quadratic dependence of $L$ on the Fermionic operator, clearly $L$ can be expressed as a quadratic polynomial in the Majorana Fermions in the following form
 \begin{equation}\label{SLDGS}
 L_{}=: \frac{1}{2}\bm{\omega}^{T}\cdot K\cdot  \bm{\omega} + \bm{\zeta}_{}^{T}\cdot \bm{\omega} +\eta,
 \end{equation}
 where $K:=\{K_{jk}\}_{jk=1}^{2n}$ is $2n\times2n$ Hermitian anti-symmetric matrix, $\bm{\zeta}:=\{\zeta^{k}\}_{k=1}^{2n}$ is a $2n$ real vector, and $\eta$ is a real number. Note that any odd-order correlation function for a Gaussian Fermionic state vanishes identically, then  
 \be
 \langle  \omega_{k}\rangle = \Tr{(\rho \omega_{k})}=0 \qquad \forall  k=1 \dots 2n\,.
 \ee 
By differentiating the above equation, one readily shows that the linear term in~(\ref{SLDGS}) is identically zero
 \[
 0=\frac{d }{d\lambda}\Tr{(\rho \omega_{k})}=\Tr{(\omega_{k} \dot \rho )}= \frac{1}{2} \Tr(\omega_{k}\{L,\rho\}) = \Tr(\rho\{\bm{\zeta}^{T}\bm{\omega},\omega_{k}\})=\zeta^{k}\,,
 \]
where $\zeta^{k}$ is the $k$-th component of $\bm{\zeta}$, and in the forth equality one takes into account that the third order correlations vanish. The quantity $\eta$ can be determined from the trace preserving condition, i.e. 
\be
0=\frac{d }{d\lambda}\,\Tr\, \rho=\Tr{(\dot {\rho})}=\Tr{(\rho L)}\,,
\ee
which, after plugging in Eq.~(\ref{SLDGS}), leads to
\begin{equation}\label{eta}
\eta=-\frac{1}{2}\Tr{(\rho \bm{\omega}^{T}K \bm{\omega})}=\frac{1}{2}\Tr{(K_{}\, \Gamma)}.
\end{equation}
 In order to determine $ K$, let's take differential of $\Gamma_{jk}=1/2\Tr{(\rho[\omega_{j},\omega_{k}])}$, then
 \begin{align}
 \dot \Gamma_{jk}=\frac{1}{2}\Tr{(\dot \rho[\omega_{j},\omega_{k}])}&=\frac{1}{4}\Tr{(\{\rho,L_{}\}[\omega_{j},\omega_{k}])}\nonumber\\
 &= \frac{1}{8}\Tr{(\{\rho,\bm{\omega}^{T} K_{}\bm{\omega}\}[\omega_{j},\omega_{k}])}+ \frac{\eta_{}}{4}\Tr{(\rho[\omega_{j},\omega_{k}])}\nonumber\\
 &=\frac{1}{8}\sum_{lm}K_{}^{lm}\Tr{(\{\rho,[\omega_{l},\omega_{m}]\}[\omega_{j},\omega_{k}])}+\frac{\eta_{}}{2} \Gamma_{jk}\nonumber\\
 &=(\Gamma  K_{} \Gamma- K_{})_{jk} + \frac{1}{2}\left[ \eta - \frac{1}{2}\Tr{( K_{}\, \Gamma)}\right]\Gamma_{jk},
 \end{align}
 where the last equality is obtained with the help of Eq.~(\ref{Pf}) and using the antisymmetry of $\Gamma$ and $  K$ under the exchange of $j$ and $k$. Finally, according to Eq.~(\ref{eta}), the last term vanishes and we obtain the following (discrete time) Lyapunov equation
\begin{equation}\label{DLE} 
\dot \Gamma = \Gamma K_{}\Gamma -K_{}.
\end{equation}
The above equation can be formally solved by
\be\label{eq:SolK}
 K_{}=(\Ad_{\Gamma}-\one)^{-1}(\dot \Gamma),
\ee
where $\Ad_{\Gamma}(X):=\Gamma X \Gamma^{\dagger}$ is the adjoint action. In the eigenbasis of $\Gamma$, (i.e. $\Gamma\ket{j}=\gamma_{j}\ket{j}$) it reads
\begin{equation}\label{Kappa}
\bk{j | K_{}|k}=( K_{})_{jk}=\frac{(\dot\Gamma)_{jk}}{\gamma_{j}\gamma_{k}-1}= -\frac{\dot \Omega_{k}}{2}\delta_{jk}+\tanh{\frac{\Omega_{j}-\Omega_{k}}{2}} \bk{j| \dot k},
\end{equation}
where, in the second equality, we made use of the relation $\gamma_{k}=\tanh{(\Omega_{k}/2)}$, which yields the following diagonal $(\dot\Gamma)_{jj}=(1-\gamma_{j}^{2})\dot\Omega_{j}$ and off-diagonal terms $(\dot\Gamma)_{jk}=(\gamma_{k}-\gamma_{j})\bk{j| \dot k}$. This expression is well defined everywhere except for $\gamma_{j}=\gamma_{k}=\pm 1$, where the Gaussian state $\rho$ becomes singular (i.e. it is not full rank). In this condition, the expression~(\ref{Kappa}) for the SLD  $L$ may become singular. Nevertheless,  the boundness of the function $|\tanh{\frac{\Omega_{j}-\Omega_{k}}{2}}|\le 1$ in~(\ref{Kappa}) shows that such a singularity is relatively benign. One can show that this condition $\gamma_{j}=\gamma_{k}=\pm 1$ produces, at most, removable singularities in the Fisher Information Matrix (cf.~\cite{Safranek2017}). This allows the quantum Fisher information matrix to be extended by continuity from the set of full-rank density matrices to the subset with $\gamma_{j}=\gamma_{k}=\pm 1$.

\section{Conclusions}
In this work we derived a general expression for the symmetric logarithmic derivative of an arbitrary Fermionic Gaussian state. We obtained a compact expression in terms of correlation matrix of a FGS, which allows for the calculation of the quantum Fisher information. This provides a way of assessing the ultimate precision of parameter estimation problems in many-body systems involving Fermionic Gaussian states, through the Cramer-Rao bound. Moreover, the expression of the SLD allows for the explicit derivation of the eigenbasis associated to the optimal quantum measurement associated to the estimation of a parameter of interest. The generality of the method offers also a way of evaluate the so called \emph{compatibility condition} in multi-parameter quantum estimation problems~\cite{Ragy2016}. Indeed, due to the quantum nature of the underlying probe, the multi-parameter estimation problem may not saturate the multi-parameter Cramer-Rao bound. Unlike classical estimation problems, the non-commutativity of the observables involved in the optimal quantum measurements associated to independent parameters may prevent the CR bound from being saturated~\cite{Szczykulska2016}. An explicit quantitative condition~\cite{Ragy2016} for such a compatibility can be easily derived, once the formula for the SLD is given. The general framework presented provides a way of easily assess the above mentioned quantities. Moreover, the explicit expression of the SLD, in analogy to Bosonic Gaussian estimation problems~\cite{Nichols2017}, can be exploited, in combination with efficient numerical algorithms, to find optimal Fermionic Gaussian probe that minimises the overall multi-parameter estimation problem~\cite{Knott2016}. 

\section*{acknowledgments}
This work was supported by the Grant of the Government of the Russian Federation (contract No. 14.Y26.31.0021). We acknowledge also partial support by Ministry of Education, University and Research of the Italian Government.


\bibliography{\bibliopath library}



\end{document}